\documentclass[letterpaper,10pt]{article} 

\usepackage{osameet3} 

\usepackage{amsmath,amssymb}
\usepackage[colorlinks=true,bookmarks=false,citecolor=blue,urlcolor=blue]{hyperref} 

\begin{document}

\title{Imaging Scatterometer for Observing Changes to Optical Coatings During Air Annealing}

\author{Michael Rezac$^1$, Daniel Martinez$^1$, Amy Gleckl$^1$, Joshua R. Smith$^{1,*}$}

\address{$^1$ The Nicholas and Lee Begovich Center for Gravitational-Wave Physics and Astronomy, California State University, Fullerton, US}
\email{$^*$josmith@fullerton.edu}

\copyrightyear{2022}

\begin{abstract}
Air annealing generally reduces absorption, scattering, and mechanical loss in amorphous coatings up to temperatures where damage occurs. Our instrument uses an industrial oven with viewports to observe coating scatter and damage during annealing.

\end{abstract}

\section{Introduction}
Optical interference coatings with as low as possible optical and mechanical losses are in demand for high-precision measurement applications. 
The gravitational-wave observatories LIGO and Virgo use coatings formed by TiO$_2$-doped Ta$_2$O$_5$ (high index) and SiO$_2$ (low index) layers, in which the dopant is added to frustrate crystallization in Ta$_2$O$_5$, allowing higher annealing temperatures~\cite{Granata:2019fye}. Post-deposition annealing of such coatings to 600$^{\circ}$C, in air, has been shown to reduce their scatter~\cite{Sayah:21,Capote:21}, absorption~\cite{Fazio:20}, and mechanical loss~\cite{Granata:2019fye}. High-temperature post-deposition annealing remains a promising path toward producing coatings with even lower mechanical loss using materials such as TiO$_2$-doped GeO$_2$ with SiO$_2$. 

The practical limits to the maximum annealing temperatures achievable for doped coatings are set by crystallization or damage mechanisms such as delamination, bubbles, and cracks. 
Traditionally, optics are inspected before and after a given annealing regimen, providing only weak insight to guide the manufacturing process. Our previous work introduced an in-situ method to observe optical scattering from coatings while they are annealed in vacuum and showed that scattered light from  TiO$_2$:Ta$_2$O$_5$, decreases during annealing in vacuum~\cite{Capote:21}. However, air annealing is likely superior to vacuum annealing in terms of stoichiometry and absorption losses. Here we describe a new instrument that was developed with the goal of imaging scattered light from the coatings while they are being annealed in air to temperatures as high as 800$^{\circ}$C. This instrument will provide deeper insight to coating damage and crystallization mechanisms by measuring their onset and evolution versus temperature. 

\section{Experimental Setup and Process}

\begin{figure}[ht]
  \centering
  \includegraphics[width=\linewidth]{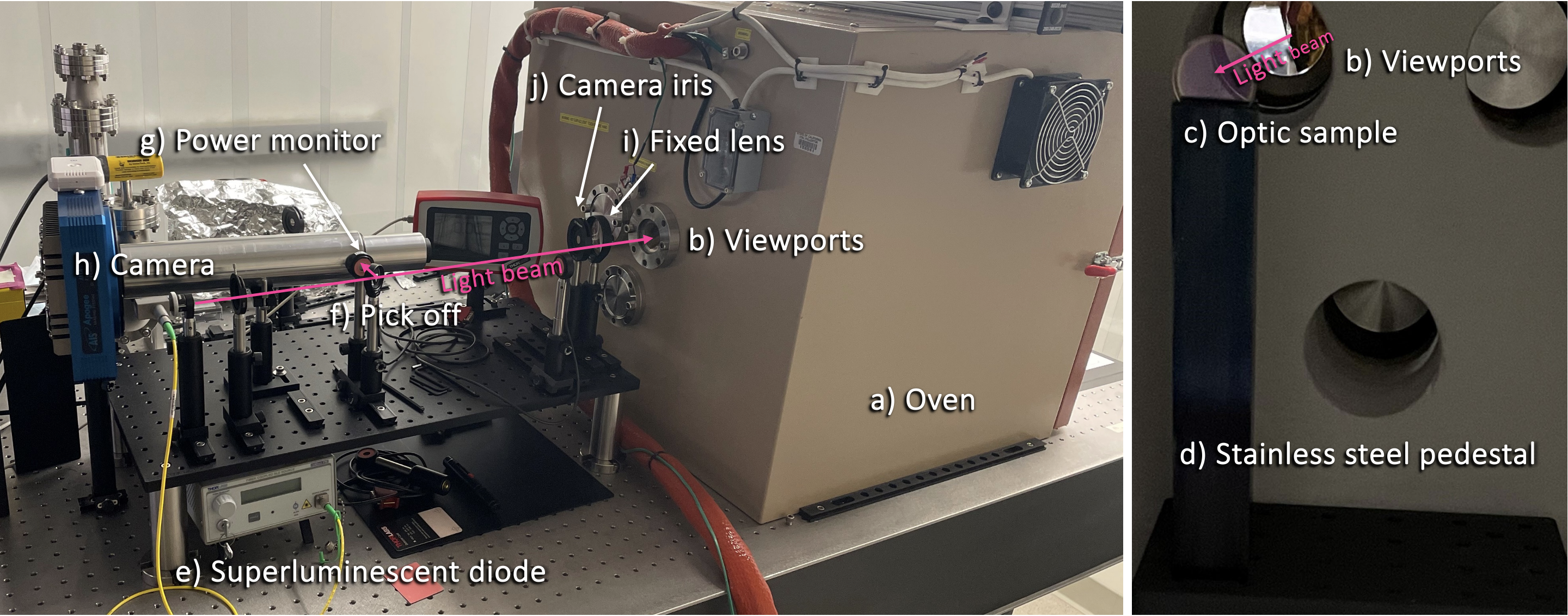}
  \caption{Setup of the Air Annealing Scatterometer. Labeled components are described in the text. \label{fig:setup}}
\end{figure}

The setup of the Air Annealing Scatterometer (AAS) is shown in Figure~\ref{fig:setup}. The basic operation is as follows. A sample coated optic is mounted within the oven, then illuminated by an external light source and imaged at regular intervals (once per minute here) using a CCD camera while an annealing temperature profile is carried out. 
This system was designed with the goal of imaging coating scatter in-situ to 800$^{\circ}$C. The components that were required to meet that goal and their interplay are described below.

The setup requires  a programmable oven (\textbf{a}) capable of reaching 800$^{\circ}$C and of accurately carrying out heating profiles with multiple ``ramp" (increase/decrease temperature) and ``dwell" (maintain constant temperature) segments.
We chose the industrial annealing oven ST-1500C-121012 by SentroTech, which uses Molybdenum disilicide (MoSi$_2$) heating elements and thick ceramic insulation to reach, for an unmodified oven, 1500$^{\circ}$C. We worked with Sentrotech's engineering team to add observation and instrument ports (\textbf{b}) to both the front (door) and rear. 
Because of these holes through the outer walls and insulation, the oven's maximum temperature will be reduced to 900$^{\circ}$C-1100$^{\circ}$C. We also added an air circulation fan option to the interior of the oven to ensure temperature consistency. The temperature is read by an S-type thermocouple that communicates with the programmable controller. 
The oven's controller (Nanodac from Eurotherm) uses proportional–integral–derivative control and provides an interface for creating heating profiles with up to 30 segments using the software package, iTools. 

A  coated sample optic (\textbf{c}) is mounted within the oven on  a solid stainless steel pedestal (\textbf{d}) so that it can be illuminated and imaged through a viewport on the back of the oven.
To provide useful information for gravitational-wave detectors, the setup requires monitoring scattered light from samples illuminated at normal incidence by a light source similar in wavelength to that used by LIGO and Virgo (1064\,nm). To avoid time-dependent speckle effects associated with coherent light (allowing association of small changes in scatter with physical changes in the coatings), we use a 1050\,nm superluminescent diode (SLD) (\textbf{e}, Thorlabs S5FC1050P). 
To monitor fluctuations in the incident power, a few percent of the SLD's output is picked off by a beam sampler (\textbf{f}, Thorlabs BSF10-C) and recorded by a  power meter (\textbf{g}, Thorlabs PM100D). The transmitted light is measured by a second power meter after passing the viewport on the front door of the oven. 
The setup could be modified to record the reflected power should in-situ measurement of reflectivity be desired. 


A low-noise and high resolution camera is required to image the light scattered from the coated optic and identify defects such as point scatterers, bubbles, and crystals. The AAS uses a cooled 4096x4096-pixel astronomical CCD camera (\textbf{h}, Apogee Alta F16M) with adjustable exposure times, programmable capture, and high linearity over a large illumination range. 
An image of the sample's surface at a scattering angle of $\theta_s=8^{\circ}$
is cast on the CCD chip, with 2X magnification, using  a single ($f$=200\,mm) lens (\textbf{i}) and adjustable iris (\textbf{j}). The SLD beams and camera are at the same height.
A narrow-band filter (Edmunds
1050nm/50nm) is installed at the front of the lens tube to limit blackbody radiation from the oven's heaters and room light from entering the camera. 
To further limit the effects of blackbody radiation and to account for ``hot" pixels in the camera, for each ``bright" image that is taken with the SLD illumination on, a ``dark" image is also taken with the SLD off, that can be subtracted from the bright image during analysis. 
Images are recorded using the Flexible Image Transport System (FITS) format, which incorporates metadata such as time stamp, exposure time, camera temperature, and saturation levels. 


A LabView Virtual Instrument (VI) is used to automatically control and acquire data from the SLD, oven, camera, and power monitors. 
A typical experiment goes as follows. A sample is installed and focus of the imaging optics is checked. The desired heating profile, with typical duration of 1-2 days, is created using iTools and loaded to the oven's controller. The VI is configured with the desired camera exposure time and imaging cadence. The VI is started and it executes the following sequence (where times indicate the duration spent in each state): i) read oven set point, heater power, and thermocouple temperature ($<$1s); ii) turn SLD on (1.5s); iii) read incident power monitor ($<$1s); iv) bright image exposure (5s); v) transfer bright image (20s); vi) turn SLD off (1.5s); vii) dark image exposure (5s); viii) transfer dark image (20s); ix) wait (roughly 9s) until the total elapsed time of the entire sequence reaches 60s, then repeat. In this way, one bright image and dark image along with one data point each for incident laser power and oven temperature data is collected per minute. The images and data are written to disk on the PC running LabView and backed up for analysis.
The setup is located on a passive seismic isolation optical bench, within a laminar flow softwall cleanroom and the room is kept closed and dark during measurement. 




At elevated temperatures, thermal radiation from the heating elements provides an alternative side illumination of the coatings that is particularly useful for viewing bubbles and delamination, especially in the ``dark" images with the SLD off. 
Results making use of this will be described in a future publication. 
We note that a dedicated side illumination could be added to the setup to achieve the same outcomes throughout the duration of the experiment.

\section{Results: crystallization of TiO$_2$:Ta$_2$O$_5$}

\begin{figure}[ht]
  \centering
  \includegraphics[width=\linewidth]{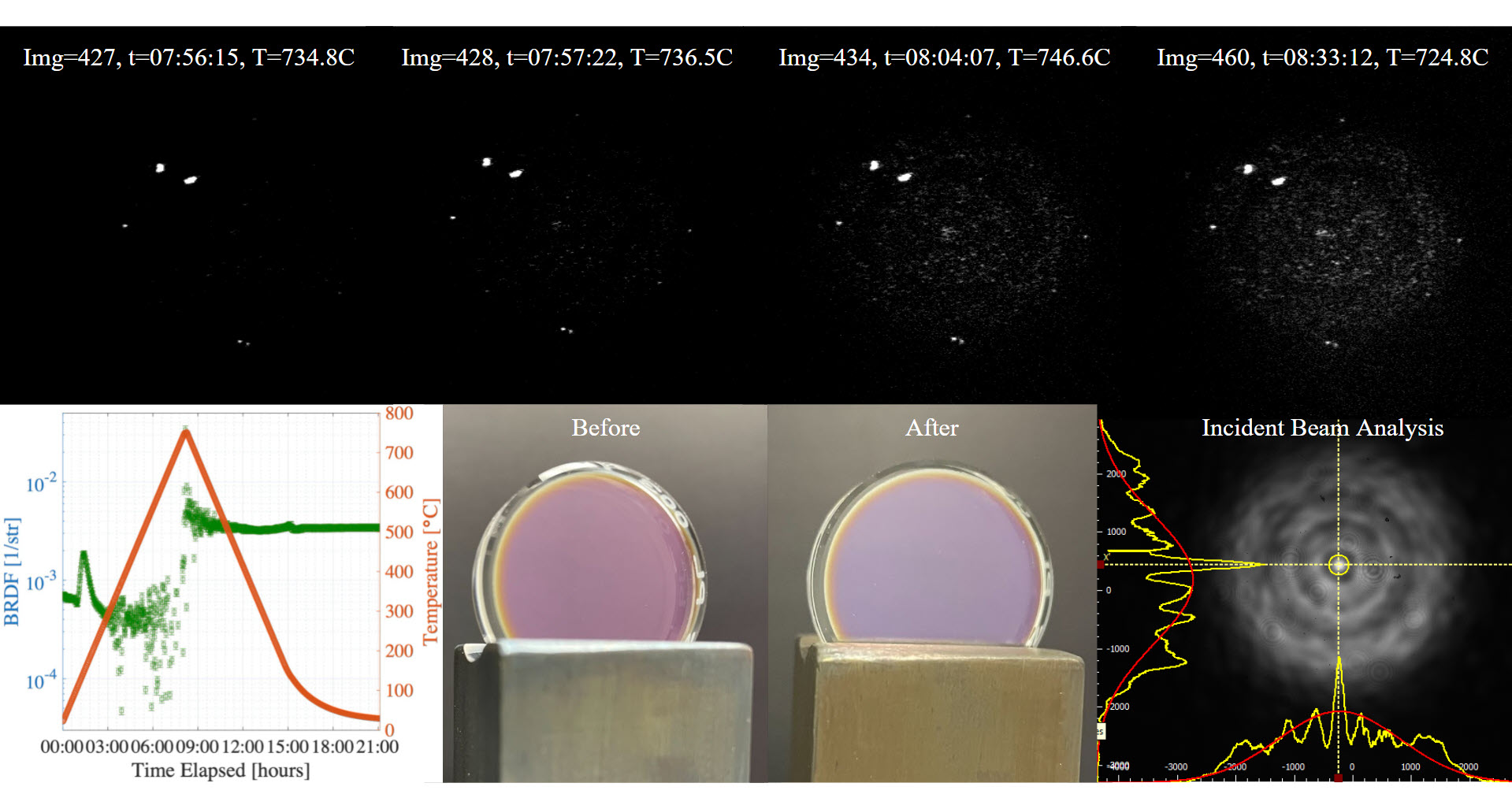}
  \caption{Results of annealing sample PL003, a single quarter-wavelength layer of TiO$_2$:Ta$_2$O$_5$ on a superpolished fused silica substrate. \textit{Bottom, left to right:} Measured oven temperature (orange) and sample BRDF (green) of the sample, both versus elapsed time; sample before annealing; sample after annealing (note, the stainless steel holder changed color); intensity profile of the incident laser beam at the location of the sample (beam diameter 4\,mm). The top row shows cropped images of the  (7\,mm-wide) region of the sample illuminated by the SLD. \textit{Top, left to right:} At 734.8$^{\circ}$C some point scatterers are visible, but no crystallization is seen; at 736.5$^{\circ}$C weak diffuse scattering from the onset of crystallization is seen; at 746.6$^{\circ}$C the beam intensity profile is scattered somewhat uniformly by the crystallized coating; at 724.8$^{\circ}$C on the ramp down, the same pattern is seen more brightly.  \label{fig:results}}
\end{figure}

Figure~\ref{fig:results} shows the results of annealing sample optic PL003, consisting of a single quarter-wavelength-thick layer of TiO$_2$:Ta$_2$O$_5$ coated (by Laboratoire des Mat\'{e}riaux Avanc\'{e}s) on a superpolished ($\sigma<0.1$\,nm) Corning 7979 fused silica substrate (from Coastline Optics), to 750$^{\circ}$C. This simple heating profile was chosen to achieve crystallization in the coating layer based on previous studies~\cite{Fazio:20}. The bottom left panel shows the time evolution of the oven's temperature and the optic's bidirectional reflectance distribution function (BRDF). The BRDF exhibits a first bright peak at low temperature due to point scatterers getting very bright as the sample moves up (due to thermal expansion of the tall stainless steel holder) and the beam sweeps over them.
At mid-temperatures the BRDF is modulated somewhat, likely also due to the sample translating within the beam pattern.
Starting around 735$^{\circ}$C there is a factor of ten, and seemingly permanent, increase in BRDF which we associated with the onset of crystallization. This increased scatter has the onset temperature and ``frosted" appearance we expect, as shown in the images along the top row, which resemble the incident beam pattern shown in the bottom right.   

\section{Conclusion}

The AAS instrument is capable of observing coating optical scatter, crystallization, and damage throughout in-air annealing to 750$^{\circ}$C, or higher. 
The images and BRDF presented here, with SLD normal illumination, reveal the onset of coating crystallization. Followup studies will be done to more closely observe crystal formation, for example heating profiles with slower temperature increases or longer soaks around the crystallization onset temperature. 
An upcoming paper will present similar onset and evolution of damage mechanisms such as bubbles and delamination seen also with side illumination from the oven's heaters. 
Together, these capabilities allow deeper study of the damage mechanisms at play in coatings during high-temperature annealing. Such studies may lead to improvements in the coating manufacture process for low-optical-loss applications.



\bibliographystyle{osajnl} 
\bibliography{references} 

\end{document}